\documentclass[aps,amssymb,prd,showpacs,twocolumn]{revtex4}
\usepackage{amsmath}
\begin{document}

\title{Einstein boundary conditions in relation to constraint propagation for
the initial-boundary value problem of the Einstein equations}

\author{Simonetta Frittelli}
\email{simo@mayu.physics.duq.edu}
\affiliation{Department of Physics, Duquesne University,
       Pittsburgh, PA 15282}
\affiliation{Department of Physics and Astronomy, University of Pittsburgh,
       Pittsburgh, PA 15260}
\author{Roberto G\'omez}
\email{gomez@psc.edu}
\affiliation{Pittsburgh Supercomputing Center, 4400 Fifth Avenue,
     Pittsburgh, PA 15213}
\affiliation{Department of Physics and Astronomy, University of Pittsburgh,
       Pittsburgh, PA 15260}

\date{\today}

\begin{abstract}

We show how the use of the normal projection of the Einstein tensor as a set of
boundary conditions relates to the propagation of the constraints, for two
representations of the  Einstein equations with vanishing shift vector: the ADM
formulation, which is ill-posed, and  the Einstein-Christoffel formulation,
which is symmetric hyperbolic.  Essentially, the components of the normal
projection of the Einstein tensor that act as non-trivial boundary conditions
are linear combinations of the evolution equations with the constraints that
are not preserved at the boundary, in both cases.  In the process, the
relationship of the normal projection of the Einstein tensor to the recently
introduced ``constraint-preserving'' boundary conditions becomes apparent.

\end{abstract}
\maketitle

\section{Introduction}\label{sec:1}

The numerical integration of the initial-boundary value problem of the vacuuum
Einstein equations is characterized, as is well known~\cite{yorksources}, by the
presence of four differential equations that must be satisfied by the solution of
the initial-value problem. These four additional equations are normally written in
the form of constraints -- that is: differential equations with no time derivatives
of the fundamental variables--, and represent the vanishing of the projection of the
Einstein tensor along the normal $n^a$ to the time foliation, namely: $G_{ab}n^b=0$.
This atypical evolution problem has been a constant source of controversy from the
onset of numerical implementations, because it is commonly found that the numerical
solution of the initial-boundary value problem with constrained initial data fails
to solve the constraints soon after the initial time. Earlier numerical simulations
predominantly opted for a scheme that forced the constraints at every time slice
(see, for instance,~\cite{evans}). In this scheme, the numerical solution of the
initial-boundary value problem at a given time step is used as a starting guess to
generate a numerical solution of the constraint equations on the same time slice,
which is subsequently used as data for the initial-boundary value problem for the
next slice, and so on.  In more recent years, this forced constrained scheme has
been abandoned in favor of unconstrained simulations that disregard the constraint
equations, except that the growth of the constraint violations is monitored and kept
to a minimum by stopping the numerical simulations before the constraint violations
grow too large (see, for instance, \cite{shibata03}).  

The reason for the generically observed drifting of numerical simulations away from
the constrained solution has yet to be pinned down and may turn out to be a
combination of analytic and computational issues. Several analytical factors have
been identified that may have a role to play in this respect, ranging from
ill-posedness to asymptotic instability of the propagation of the
constraints~\cite{frittellinote,lambdavar}. In the face of the recurrence of
uncontrolled numerical instabilities that prevent numerical simulations from
maintaining accuracy for a time scale commensurate with the emission of strong
gravitational waves from gravitational collapse, the importance of constraint
violations in numerical simulations had widely been regarded as secondary until
recently.  In \cite{teukolsky}, the authors show, with sufficiently accurate numerical
experiments, that the growth of constraint violations plays a key role in shortening
the running time of numerical simulations even when other factors are kept within
control.  Understanding the growth of constraint violations and figuring out
techniques to keep them in check have become issues of high priority in the goal to
extend the run time of numerical simulations to a regime where gravitational
radiation can be extracted. 
  
What we show in this article is that the projection of the Einstein tensor
along the normal $e^a$ to the boundary of the simulation, $G_{ab}e^b$, is
equivalent to the constraints that are not identically satisfied at the
boundary by virtue of the initial-value problem alone. This is probably a
generic fact, irrespective of the details of the formulation of the
initial-boundary value problem of the 3+1 Einstein equations. We prove this
fact in the case of the ADM formulation, which is ill posed but has well-posed
constraint propagation, and also in the case of the Einstein-Christoffel
formulation, which is well posed and has well-posed constraint propagation. 

This fact connects intimately two factors of seemingly independent effect to
numerical simulations; i.e., the problem of constraint violations and the problem of
proper boundary conditions.  Intuitively, if boundary values travel inwardly at the
speed of light (or any finite speed), then their effect is confined to the sector of
the final time slice that lies in the causal future of the boundary surface,
regardless of the extent of the bulk. Thus if the bulk is significantly larger than
the run time in units of the speed of light, the boundary effects will be negligible
\textit{in the bulk.\/} As a result, it is commonly assumed that boundary effects
can normally be kept in check by ``pushing the boundaries out'' and this has largely
become the implicit rule in numerical simulations. (This is mathematically true only
in the case of strongly hyperbolic formulations, though.) The constraint violations,
on the other hand, appear to remain in most cases. The two problems would thus seem
independent of each other (in fact, they are itemized separately in
\cite{teukolsky}).

On closer inspection, given any formulation of the initial-boundary value
problem there are constraints that are identically satisfied in the simulated
region by virtue of a proper choice of initial values, and there is a subset of
constraints that are not satisfied in the entire simulated region by virtue of
the initial values alone. In a minimallistic sense, it is these ``nonpreserved''
constraints that need to be controlled. Because, as we show here, these
``nonpreserved'' constraints are equivalent (up to evolution) to $G_{ab}e^b=0$,
and because $G_{ab}e^b=0$, in turn, are proper and necessary boundary
conditions for the initial-boundary value problem--as we have shown previously
\cite{boundary3d}--, it follows that by treating the boundaries properly one
may get a handle on keeping the relevant constraint violations in check as a
byproduct -- a reasonable deal, by all means. 

The article is organized as follows. The ADM case, which is the simplest and
most transparent one in spite of its modest repertory of useful mathematical
attributes, is dealt with first in Section~\ref{sec:2}.  As an example of
strongly hyperbolic formulations, the Einstein-Christoffel case is developed in
Sections~\ref{sec:3} and \ref{sec:4}. Both cases are treated with the
simplifying assumption of vanishing shift vector in the terminology of the 3+1
split. The reader may assume that the complexity of the argument would increase
sharply in the presence of a non-vanishing shift vector, although its general
gist should run unchanged. This work develops issues that were anticipated in
\cite{boundary3d}.

\section{ADM formulation with vanishing shift\label{sec:2}}

Throughout the article we assume the following form for the metric of spacetime
in coordinates $x^a = (x^i,t)$ in terms of the three-metric $\gamma_{ij}$ of
the slices at fixed value of $t$:
\begin{equation}\label{metric}
    ds^2 = -\alpha^2 dt^2 + \gamma_{ij}dx^idx^j
\end{equation}

\noindent where $\alpha$ is the lapse function. The Einstein equations
$G_{ab}=0$ for the four-dimensional metric are equivalently expressed in the
ADM form~\cite{yorksources}:
\begin{subequations}\label{adm}
\begin{eqnarray}
    \dot{\gamma}_{ij} &=& - 2\alpha K_{ij}, \label{adma}\\
    \dot{K}_{ij} &=& \alpha \left(R_{ij} - 2 K_{il}K^l{}_j
            +K K_{ij}\right) -D_iD_j\alpha, \label{admb}
\end{eqnarray}
\end{subequations}

\noindent with the constraints
\begin{subequations}\label{admconst}
\begin{eqnarray}
    {\cal C} &\equiv &-\frac12\left(R - K_{ij}K^{ij} + K^2 \right)= 0,   \\
    {\cal C}_i&\equiv &D_jK^j{}_i-D_i K = 0.
\end{eqnarray}
\end{subequations}

\noindent Here an overdot denotes a partial derivative with respect to the time
coordinate ($\partial/\partial t$), indices are raised with the inverse metric
$\gamma^{ij}$, $D_i$ is the covariant three-derivative consistent with
$\gamma_{ij}$ , $R_{ij}$ is the Ricci curvature tensor of $\gamma_{ij}$, $R$
its Ricci scalar, $K_{ij}$ is the extrinsic curvature of the slice at fixed
value of $t$ and $K\equiv \gamma^{ij}K_{ij}$.  Expressed in terms of the
Einstein tensor, the constraints (\ref{admconst})
are related to specific components in the coordinates $(x^i,t)$:
\begin{subequations}\label{constG}
\begin{eqnarray}
	{\cal C}  &=& -\alpha^2 G^{tt}\label{constGa}\\
	{\cal C}_i&=& -\alpha \gamma_{ij}G^{jt}	,\label{constGb}
\end{eqnarray}
\end{subequations}

\noindent where (\ref{constGb}) holds only for vanishing shift vector. The
constraint character (the absence of second derivatives with respect to time)
is a consequence of the fact that the only components of the Einstein tensor that
appear in (\ref{constG}) have a contravariant index of value $t$. In geometric
terms, (\ref{constG}) are linear combinations of $G_{ab}n^b=0$ where $n^b$ is
the unit normal to the slices of fixed value of $t$, and is therefore given by
$n^a=g^{ab}n_b=-\alpha g^{at}=\delta^a_t/\alpha$.

Similarly, at any boundary given by a fixed value of a spatial coordinate, the
normal vector to the boundary surface $e^b$ can be used to project the Einstein
tensor as $G_{ab}e^b$ in order to obtain the components that have no second
derivatives across the boundary. The vanishing of these can then be imposed as
conditions on the boundary values for the fundamental variables. To fix ideas
let's choose a boundary surface at a fixed value of $x$. Thus $e^b= g^{bx}=
(0,\gamma^{ix})$ up to scaling. Consequently, $G_{ab}e^b = G_a^x$, so any
linear combination of the components of the Einstein tensor with a
contravariant index of value $x$ will be suitable. Explicitly we have:
\begin{subequations}\label{bcs}
\begin{eqnarray}
G_t^x &=& -\frac12 \gamma^{ix}\left((\ln \gamma),_{it}
          -\gamma^{kl}\dot{\gamma}_{ik,l}\right) 
-KD^x\alpha +K^x_kD^k\alpha\nonumber\\ &&
         + \alpha\left( \gamma^{kl}\Gamma^j{}_{kl}K^x_j
           +\gamma^{ix}\Gamma^j{}_{ik}K^k_j\right) \label{det}\\
G_y^x &=& -\frac{\dot{K}^x_y}{\alpha} + KK^x_y
          + R^x_y -\frac{1}{\alpha}
        D^xD_y\alpha \label{Kxy}\\
G_z^x &=& -\frac{\dot{K}^x_z}{\alpha} + KK^x_z
          + R^x_z -\frac{1}{\alpha}
        D^xD_z\alpha\label{Kxz}\\
G_x^x&=& \frac{\dot{K}-\dot{K}^x_x}{\alpha}
-\frac12 (R +K^{ij}K_{ij} +K^2) +KK^x_x\nonumber\\
&&
         +R^x_x+\frac{1}{\alpha}\left(D^jD_j\alpha- D^xD_x\alpha\right)
\label{K-Kxx}
\end{eqnarray}
\end{subequations}

\noindent  Here
$\Gamma^k{}_{ij}=(1/2)\gamma^{kl}(\gamma_{il,j}+\gamma_{jl,i}-\gamma_{ij,l})$,
and the time derivative of the components of the extrinsic
curvature is applied after raising an index, that is: $\dot{K}^i_j
\equiv (\gamma^{ik}K_{kj}),_t$. The reader can verify that $R^x_y$ and
$R^x_z$ do not involve second derivatives with respect to $x$ of any
of the variables and that the combination $R^x_x -\frac12 R$ doesn't
either.

At this point the question arises as to whether $G_{ab}e^b$ vanish identically
on the boundary for any solution of the evolution equations (\ref{adm}) with
initial data satisfying (\ref{admconst}). To start with, by inspection one can
clearly see that $G_y^x$ and $G_z^x$ are exactly the evolution equations
(\ref{admconst}) for the mixed components $K_y^x$ and $K_z^x$ of the extrinsic
curvature (namely: Eq.~(\ref{admb}) multiplied by $\gamma^{ix}$ and evaluated
at $j=y$ and $j=z$). Thus two of the four boundary equations are identically
satisfied by the solution of the evolution equations (irrespective of the
initial data).  That is not the case with $G_x^x$. If one uses the evolution
equations for $K_x^x$ and for the trace $K$ that follow from (\ref{admb}) to
eliminate the time derivatives from $G_x^x$, one is left with the Hamiltonian
constraint ${\cal C}$.  Finally,  using the definition of the extrinsic
curvature (\ref{adma}) into the expression for $G_t^x$ it is straightforward to
see that $G_t^x= \alpha \gamma^{xi}{\cal C}_i$ (this is necessarily so in the
case of vanishing shift because $G_t^x = g_{ta}G^{ax} = -\alpha^2 G^{tx}$ and
inverting (\ref{constGb}) one has that  $G^{tx}= -\alpha^{-1}\gamma^{xi}{\cal
C}_i$). In summary, if we represent the evolution equations (\ref{adma}) and
(\ref{admb}) in the form ${\cal E}^\gamma_{ij}=0$ and ${\cal E}^K_{ij}=0$
respectively by transferring all the terms from the right into the left-hand
side, we have
\begin{subequations}\label{bctoconst}
\begin{eqnarray}
G_t^x &=& \alpha{\cal C}^x 
	+ \frac12\gamma^{xj}\gamma^{kl}(\partial_l{\cal E}^{\gamma}_{jk}
	- \gamma^{xl}\gamma^{jk}\partial_l{\cal E}^{\gamma}_{jk})\label{bctoconsta}\\
G_y^x &= & \frac{1}{\alpha}\gamma^{xj}{\cal E}^{K}_{jy} \label{bctoconstb}\\
G_z^x &=& \frac{1}{\alpha}\gamma^{xj}{\cal E}^{K}_{jz}\label{bctoconstc}\\
G_x^x &= & -{\cal C} +\frac{1}{\alpha}(\gamma^{kl}{\cal E}^K_{kl} - \gamma^{xj}{\cal
E}^K_{xj})\label{bctoconstd}
\end{eqnarray}
\end{subequations}

\noindent which can essentially be expressed in the form 
\begin{subequations}\label{equivbctoconst}
\begin{eqnarray}
G_t^x &\sim& \alpha{\cal C}^x \label{equivbctoconsta}\\
G_y^x &\sim & 0 \label{equivbctoconstb}\\
G_z^x &\sim & 0\label{equivbctoconstc}\\
G_x^x &\sim & -{\cal C} \label{equivbctoconstd}
\end{eqnarray}
\end{subequations}

\noindent where the symbol $\sim$ represents \textit{equality up to terms
proportional to the evolution equations.\/}  Clearly the vanishing of $G_t^x$
and $G_x^x$ at the boundary is intimately connected with the propagation of 
${\cal C}$ and ${\cal C}^x$ towards the boundary. On the other hand,  ${\cal
C}^y$ and ${\cal C}^z$ are not related to the boundary equations.  

To have a sense for why this should be so, one may now look at the auxiliary
system of propagation equations for the constraints as implied from
(\ref{adm}). Taking a time derivative of the constraints and using the
evolution equations we have
\begin{subequations}\label{constprop}
\begin{eqnarray}
\dot{\cal C}  &=& \alpha \partial^i {\cal C}_i + \ldots \\
\dot{\cal C}_i &=& \alpha \partial_i{\cal C} + \ldots
\end{eqnarray}
\end{subequations}  

\noindent where $\ldots$ denote undifferentiated terms. This is a well-posed
system of equations in the sense that it is \textit{strongly
hyperbolic}~\cite{kreissbook}. The characteristic speeds are $0$ and $\pm
\alpha$. With respect to the unit vector $\xi^i =
\gamma^{ix}/\sqrt{\gamma^{xx}}$, normal to the boundary, the characteristic
fields that travel with zero speed are ${\cal C}^y$ and ${\cal C}^z$. This is
an indication that these constraints could be interpreted as ``static'', and
that  their vanishing is preserved by the evolution at all times after the
initial slice, that is: they will vanish at the boundary by virtue of the
evolution equations and the initial values (of course, only so long as the
other constraints also vanish, due to the coupling in the undifferentiated
terms in the constraint propagation equations). It makes sense, thus, that they
do not relate at all with the boundary equations $G_a^x=0$ and that two of the
boundary equations are redundant (i.e., trivial). 

The characteristic fields that travel with non-zero speeds $\pm\alpha$ are,
respectively, ${\cal C}^\pm\equiv {\cal C} \pm {\cal C}^x/\sqrt{\gamma^{xx}}$. 
The characteristic field ${\cal C} + {\cal C}^x/\sqrt{\gamma^{xx}}$ is outgoing
at the boundary, which means that its value is propagated from the initial
slice to the boundary. If this constraint combination vanishes initially, then
it should also vanish at the boundary.  On the other hand, the characteristic
field ${\cal C} - {\cal C}^x/\sqrt{\gamma^{xx}}$ is incoming at the boundary,
which means that its value at the boundary is \textit{unrelated} to the initial
values prescribed in the interior. Thus, this combination of constraints is not
vanishing at the boundary by virtue of the initial data and the evolution. This
is a ``non-preserved'' constraint. This constraint must be enforced at the
boundary by means of appropriate boundary conditions if the solution of the ADM
equations (\ref{adm}) at the final time slice is to satisfy all four of the
constraints. Since the two boundary equations $G^x_x=0$ and $G^x_t=0$ are
directly related to the two characteristic constraints with non-vanishing
speeds, imposing either one or any linear combination of them \textit{except}
$G^x_t+(\alpha/\sqrt{\gamma^{xx}}) G^x_x=0$ on the boundary is equivalent to
imposing a linear relationship ${\cal C}^- = B {\cal C}^+$ (whith a constant
$B$) as a boundary condition on the system of equations for the constraints. 
These are valid boundary conditions that preserve the well-posedness of the
evolution of the constraints~\cite{kreissbook}.

Thus, of the two equations $G^x_x=0$ and $G^x_t=0$, one of them is necessary to
avoid constraint violations. The remaining one, which we can represent by
the linear combination $G^x_t+(\alpha/\sqrt{\gamma^{xx}}) G^x_x=0$, is
equivalent to a condition on the \textit{outgoing constraint} ${\cal C}^+$. 
Therefore, it is redundant to the system of evolution of the constraints.  At
this point, however, it is not clear whether this remaining equation would be
redundant as a boundary condition for the ADM evolution. The reason is that
since the ADM equations are not strongly hyperbolic, the number of boundary
conditions it requires is not known.  

We are led to infer that some of the components of the projection of the
Einstein tensor normally to the boundary are in direct correspondence with
those constraints that are not automatically preserved by the evolution
equations and the initial values. In this particular case of the ADM
formulation, because the evolution equations thenselves are not strongly
hyperbolic, perhaps nothing else can be said about the role of the equations
$G_{ab}e^b=0$ as boundary conditions, except that failure to impose the
nontrivial one leads to constraint violations with guaranteed certainty.  

Yet the case of the ADM equations has been sufficiently direct and transparent
to provide us with a guide for the much more complicated (but also much
stronger) case of formulations of the Einstein equations that are strongly
hyperbolic.  In the next Section, a particular formulation is chosen for its
simplicity and its close relationship to the ADM case.

\section{Boundary conditions for the EC formulation with vanishing shift\label{sec:3}}

Most hyperbolic formulations of the Einstein equations require the introduction
of additional variables in order to reduce the differential order in the
spacelike coordinates from second to first. The additional variables are always
a complete set of linearly independent combinations of the space-derivatives of
$\gamma_{ij}$.  This inevitably introduces an ambiguity: whereas indices
labeling second-order derivatives are automatically symmetric (which affords
plenty of convenient cancellations), the same terms written in terms of the new
variables are not manifestly symmetric, the symmetry becoming a commodity that
may or not be ``turned on'' by imposing yet additional constraints. For
example, in second order one has, unambiguously, $\gamma_{ij,kl}-\gamma_{ij,lk}
= 0$. However, if one defines $d_{ijk}\equiv\gamma_{ij,k}$, one is free to
write the same expression either as $0$ or as $d_{ijk,l} -d_{ijl,k}$. 
Undoubtedly, it makes a big difference to a numerical code to write it one way
or the other, since they are \textit{not\/} the same when it comes to the
solution-generating process. This ambiguity is usually taken advantage of in
order to write the Einstein equations in manifestly well-posed form.  The cost
is the loss of some transparency and the addition of constraints and, as we
will see shortly, boundary conditions.     

Consider the Einstein-Christoffel [EC]
formulation of the 3+1 equations~\cite{fixing}.  By defining
first-order variables as the following 18 linearly independent
combinations of the space-derivatives of the metric:
\begin{equation}\label{fkij}
f_{kij} \equiv \Gamma_{(ij)k} +
\gamma_{ki}\gamma^{lm}\Gamma_{[lj]m}
+\gamma_{kj}\gamma^{lm}\Gamma_{[li]m},
\end{equation}

\noindent and choosing the lapse function as $\alpha\equiv Q
\sqrt\gamma$ with $Q$ assumed arbitrarily prescribed a priori, the
3+1 equations can be put in the following equivalent
form~\cite{teukolsky}:
\begin{subequations}
\begin{widetext}
\begin{eqnarray}
\dot{\gamma}_{ij} &=& -2\alpha K_{ij}           \\
\dot{K}_{ij}+\alpha\gamma^{kl}\partial_lf_{kij} &=& \alpha \{
\gamma^{kl}(K_{kl}K_{ij} - 2K_{ki}K_{lj})
+\gamma^{kl}\gamma^{mn}(4f_{kmi}f_{[ln]j} +4f_{km[n}f_{l]ij}
-f_{ikm}f_{jln} \nonumber\\
&& +8f_{(ij)k}f_{[ln]m} +4f_{km(i}f_{j)ln} -8f_{kli}f_{mnj}
+20f_{kl(i}f_{j)mn} -13f_{ikl}f_{jmn}) \nonumber\\
&& -\partial_i\partial_j\ln Q -\partial_i\ln Q\partial_j\ln Q
+2\gamma_{ij}\gamma^{kl}\gamma^{mn}(f_{kmn}\partial_l\ln Q
-f_{kml}\partial_n\ln Q)\nonumber\\
&& +\gamma^{kl} [(2f_{(ij)k}-f_{kij})\partial_l\ln Q
+4f_{kl(i}\partial_{j)}\ln Q -3(f_{ikl}\partial_j\ln Q
+f_{jkl}\partial_i\ln Q)]\}                      \label{KdotEC}\\
\dot{f}_{kij} + \alpha\partial_k K_{ij} &=& \alpha \{
\gamma^{mn}[4K_{k(i}f_{j)mn} -4f_{mn(i}K_{j)k}
+K_{ij}(2f_{mnk}-3f_{kmn})] \nonumber\\
&& +2\gamma^{mn}\gamma^{pq}[K_{mp}(\gamma_{k(i}f_{j)qn}
-2f_{qn(i}\gamma_{j)k}) +\gamma_{k(i}K_{j)m}(8f_{npq}-6f_{pqn})
\nonumber\\
&& +K_{mn}(4f_{pq(i}\gamma_{j)k}-5\gamma_{k(i}f_{j)pq})]
-K_{ij}\partial_k\ln Q \nonumber\\
&& +2\gamma^{mn}(K_{m(i}\gamma_{j)k}\partial_n\ln Q
-K_{mn}\gamma_{k(i}\partial_{j)}\ln Q)\} \label{fdotEC}
\end{eqnarray}
\end{widetext}
\end{subequations}

\noindent with the constraints
\begin{subequations}
\begin{eqnarray}
{\cal C} &\equiv&-\frac12  \gamma^{ij}\gamma^{kl}\{
2(\partial_kf_{ijl}-\partial_if_{jkl}) +K_{ik}K_{jl}
-K_{ij}K_{kl}\nonumber\\
&& +\gamma^{mn}[f_{ikm}(5f_{jln}-6f_{ljn}) + 13f_{ikl}f_{jmn}
\nonumber\\
&& +f_{ijk}(8f_{mln}-20f_{lmn}]\} = 0 \label{CEC}  
\\    
{\cal C}_i &\equiv& -\gamma^{kl}\{
\gamma^{mn}[K_{ik}(3f_{lmn}-2f_{mnl})
-K_{km}f_{iln}] \nonumber\\
&&+\partial_iK_{kl}-\partial_kK_{il}\} =0   \\
{\cal C}_{kij} &\equiv& 2f_{kij}
-4\gamma^{lm}(f_{lm(i}\gamma_{j)k} -\gamma_{k(i}f_{j)lm})-\partial_k\gamma_{ij}
=0\nonumber\\
&&
\end{eqnarray}
\end{subequations}

\noindent to be imposed only on the initial data. Here ${\cal C}_{ijk}$
represent the definition of the additional 18 first-order variables $f_{kij}$
needed to reduce the ADM equations to full first-order form (they represent
(\ref{fkij}) inverted for $\gamma_{ij,k}$ in terms of $f_{kij}$). No other
constraints are needed to pick a solution of the Einstein equations out of the
larger set of solutions of the EC equations. A point that
deserves to be made here is that (\ref{KdotEC}) is an exact transcription of
the ADM evolution equation (\ref{admb}), with no mixing of constraints into it.
In other words: Eq.~(\ref{KdotEC}) reduces exactly to Eq.~(\ref{admb}) if 
$f_{kij}$ is substituted back in terms of $\gamma_{ij}$ (assuming the same
lapse condition). 

We need to translate the boundary equations (\ref{bcs}) in terms of the EC
variables. Because of the ambiguity in writing second derivatives
$\gamma_{ij,kl}= \gamma_{ij,lk} = (\gamma_{ij,kl}+\gamma_{ij,lk})/2$ in terms
of first derivatives of $f_{kij}$ there are, in principle, many different ways
to write $G_a^x=0$ in terms of $f_{kij}$. Notwithstanding, among all these
different possibilities, the admissible boundary equations are
determined by the requirement that no $x-$derivatives of any variables may
occur. In the following, we write the simplest expression of the boundary
equations (\ref{bcs}) up to undifferentiated terms.

We start with Eq.~(\ref{Kxy}). The terms with second derivatives arise from the
combination $R^x_y - \gamma^{xk}\alpha,_{yk}/\alpha$ where $\alpha= Q
\sqrt\gamma$ with arbitrary $Q$. Up to undifferentiated terms in the fundamental
variables this combination is
\begin{eqnarray}
R^x_y - \frac{\gamma^{xk}\alpha,_{yk}}{\alpha} &=& 
- \frac12 \gamma^{xm}\gamma^{kl}
(\gamma_{ym,kl}- \gamma_{yl,km}-\gamma_{km,yl})\nonumber\\ &&
		-\gamma^{xm}\gamma^{kl}\gamma_{kl,ym} +\ldots
\end{eqnarray}
				
\noindent which can equivalently be represented as
\begin{eqnarray}\label{mainGxy}
R^x_y - \frac{\gamma^{xk}\alpha,_{yk}}{\alpha} &=& 
- \frac12 \gamma^{xm}\gamma^{kl}
(\gamma_{ym,kl}- \gamma_{yl,km})\nonumber\\ &&
+\frac12 \gamma^{xm}\gamma^{kl} \partial_y\gamma_{km,l}\nonumber\\ &&
		-\gamma^{xm}\gamma^{kl}\partial_y\gamma_{kl,m} +\ldots
\end{eqnarray}

\noindent where it is clear that in the last two terms $\gamma_{km,l}$ and 
$\gamma_{kl,m}$ can be substituted now directly in terms of $f_{kij}$ via 
\begin{equation}\label{subs}
\gamma_{ij,k}= 2f_{kij}
-4\gamma^{lm}(f_{lm(i}\gamma_{j)k} -\gamma_{k(i}f_{j)lm})
\end{equation}

\noindent yielding 
\begin{eqnarray}
\frac12\gamma^{xm}\gamma^{kl} \partial_y\gamma_{km,l}&=& 
\partial_y(-3 f^k{}_k{}^x + 4 f^x{}_k{}^k) +\ldots\\
\gamma^{xm}\gamma^{kl}\partial_y\gamma_{kl,m} &=&
\partial_y(-4 f^k{}_k{}^x +6 f^x{}_k{}^k) +\ldots
\end{eqnarray}

\noindent Thus the last two terms clearly have no $x-$derivatives. By a
relabeling of the dummy indices the remaining term is equivalent to 
\begin{equation}
\gamma^{xm}\gamma^{kl}(\gamma_{ym,kl}- \gamma_{yl,km})
= (\gamma^{xm}\gamma^{kl}-\gamma^{xl}\gamma^{km})\gamma_{ym,kl}
\end{equation}

\noindent One can see by inspection that
the contribution of $k=x$ is identically vanishing, so we have, equivalently:
\begin{eqnarray}
(\gamma^{xm}\gamma^{kl}-\gamma^{xl}\gamma^{km})\gamma_{ym,kl} &=&
(\gamma^{xm}\gamma^{yl}-\gamma^{xl}\gamma^{ym})\gamma_{ym,yl}\nonumber\\&&+
(\gamma^{xm}\gamma^{zl}-\gamma^{xl}\gamma^{zm})\gamma_{ym,zl}\nonumber\\&=&
\partial_y
\left[(\gamma^{xm}\gamma^{yl}-\gamma^{xl}\gamma^{ym})\gamma_{ym,l}\right]\nonumber\\&&+
\partial_z\left[(\gamma^{xm}\gamma^{zl}-\gamma^{xl}\gamma^{zm})\gamma_{ym,l}\right]
\nonumber\\&&+\ldots
\end{eqnarray}

\noindent Substituting $\gamma_{ym,l}$ in terms of $f_{kij}$ via (\ref{subs}) we
have
\begin{eqnarray}
\frac12(\gamma^{xm}\gamma^{yl}-\gamma^{xl}\gamma^{ym})\gamma_{ym,l}&=&
f^y{}_y{}^x-f^x{}_y{}^y \nonumber\\ &&
- f^k{}_k{}^x+f^x{}_k{}^k +\ldots\;\;\;\\
\frac12(\gamma^{xm}\gamma^{zl}-\gamma^{xl}\gamma^{zm})\gamma_{ym,l}&=&
f^z{}_y{}^x-f^x{}_y{}^z +\ldots
\end{eqnarray}

\noindent Using all these in (\ref{mainGxy}) we finally have
\begin{eqnarray}\label{mainGxyifinal}
R^x_y - \frac{\gamma^{xk}\alpha,_{yk}}{\alpha} &=&
\partial_z\left(f^x{}_y{}^z -f^z{}_y{}^x\right) \nonumber\\&&
+\partial_y\left( f^x{}_y{}^y -f^y{}_y{}^x + 3f^x{}_k{}^k -2f^k{}_k{}^x\right)
\nonumber\\&&
 +\ldots
\end{eqnarray}
 
\noindent The principal terms in $R^x_z - \gamma^{xk}\alpha,_{zk}/\alpha$ can,
evidently, be handled in the same manner, so we have the following expressions
for (\ref{Kxy}) and (\ref{Kxz}) explicit up to undifferentiated terms in the
fundamental variables:
\begin{eqnarray}
G^x_y &=& -\frac{\dot{K}^x_y}{\alpha} +
\partial_z\left(f^x{}_y{}^z -f^z{}_y{}^x\right) \nonumber\\&&
+\partial_y\left( f^x{}_y{}^y -f^y{}_y{}^x - 3f^x{}_k{}^k +2f^k{}_k{}^x\right)
 +\ldots\;\;\;\label{bcyEC}\\
G^x_z &=& -\frac{\dot{K}^x_z}{\alpha} +
\partial_y\left(f^x{}_z{}^y -f^y{}_z{}^x\right) \nonumber\\&&
+\partial_z\left( f^x{}_z{}^z -f^z{}_z{}^x - 3f^x{}_k{}^k +2f^k{}_k{}^x\right)
 +\ldots\label{bczEC}
\end{eqnarray}

The calculation of the representation of $G^x_x$ in terms of $f_{kij}$ is
much longer, but equally straightforward. In the following only the main steps
are indicated. We start with the explicit
expression of the Ricci components in terms of the metric: 
\begin{eqnarray}
R^x_x-\frac12 R &=& 
\frac12 (\gamma^{xm}\gamma^{kl} -
\gamma^{xl}\gamma^{km})\gamma_{km,xl}\nonumber\\&&
-\frac12 (\gamma^{yl}\gamma^{km} -
\gamma^{ym}\gamma^{kl})\gamma_{ym,kl}\nonumber\\&&
-\frac12 (\gamma^{zl}\gamma^{km} -
\gamma^{zm}\gamma^{kl})\gamma_{zm,kl} +\ldots
\end{eqnarray}

\noindent Expanding appropriately some of the contractions indicated (with the
guidance that no second $x-$derivatives may remain) leads to
a large number of cancellations, yielding the equivalent form:
\begin{eqnarray}
R^x_x-\frac12 R &=& 
-\frac12 (\gamma^{zm}\gamma^{yl} -
\gamma^{zl}\gamma^{ym})\gamma_{ym,zl}\nonumber\\&&
-\frac12 (\gamma^{ym}\gamma^{zl} -
\gamma^{yl}\gamma^{zm})\gamma_{zm,yl} +\ldots
\end{eqnarray}

\noindent where it is obvious that $\partial_z$ and $\partial_y$ can be pulled
out of terms that are expressible in terms of $f_{kij}$.  We have
\begin{eqnarray}
\frac12 (\gamma^{zm}\gamma^{yl} -\gamma^{zl}\gamma^{ym})\gamma_{ym,l}&=& 
f^y{}_y{}^z - f^z{}_y{}^y \nonumber\\&&
+ f^z{}_k{}^k - f^k{}_k{}^z +\ldots\;\;\;\\
\frac12 (\gamma^{ym}\gamma^{zl} -\gamma^{yl}\gamma^{zm})\gamma_{zm,l}&=& 
f^z{}_z{}^y - f^y{}_z{}^z \nonumber\\&&
+ f^y{}_k{}^k - f^k{}_k{}^y +\ldots
\end{eqnarray} 

\noindent Consequently we have
\begin{eqnarray}\label{riccixxEC}
R^x_x-\frac12 R &=& 
\partial_z(f^z{}_y{}^y-f^y{}_y{}^z 
+ f^k{}_k{}^z-f^z{}_k{}^k)\nonumber\\&&
+\partial_y(f^y{}_z{}^z-f^z{}_z{}^y 
+ f^k{}_k{}^y-f^y{}_k{}^k ) 
+\ldots\;\;\;\;\;\;
\end{eqnarray}

The terms on second derivatives of the lapse in (\ref{K-Kxx}) must be
represented in terms of $f_{kij}$ as well. They are:
\begin{eqnarray}\label{lapsexxEC}
\frac{1}{\alpha}(D^kD_k\alpha - D^xD_x\alpha)&=&
\gamma^{ym}\frac{\alpha,_{my}}{\alpha} 
+\gamma^{zm}\frac{\alpha,_{mz}}{\alpha}+\ldots\nonumber\\
&=&  \frac12 \gamma^{kl}\left(\gamma^{ym}\gamma_{kl,my}
	+\gamma^{zm}\gamma_{kl,mz}\right)\nonumber\\&& 
+\ldots\nonumber\\
&=& \partial_y\left(3f^y{}_k{}^k -2f^k{}_k{}^y\right)\nonumber\\&&
   +\partial_z\left(3f^z{}_k{}^k -2f^k{}_k{}^z\right) \nonumber\\&&
+\ldots
\end{eqnarray}

With (\ref{riccixxEC}) and (\ref{lapsexxEC}), the boundary equation (\ref{K-Kxx})
reads
\begin{eqnarray}\label{bcxEC}
G_x^x&=& \frac{\dot{K}-\dot{K}^x_x}{\alpha} \nonumber\\&&
+\partial_z(f^z{}_z{}^z+2f^z{}_x{}^x - f^x{}_x^z +3f^z{}_y{}^y
-2f^y{}_y{}^z)\nonumber\\ &&
+\partial_y(f^y{}_y{}^y+2f^y{}_x{}^x - f^x{}_x^y +3f^y{}_z{}^z
-2f^z{}_z{}^y) \nonumber\\&&
+\ldots
\end{eqnarray}

Finally, it is quite straightforward to translate the boundary equation
(\ref{det}) in terms of $f_{kij}$:
\begin{equation}\label{bctEC}
G^x_t = \dot{f}^x{}_k{}^k - \dot{f}^k{}_k{}^x +\ldots
\end{equation}

\noindent where $\dot{f}^k{}_i{}^j\equiv
\partial_t(\gamma^{km}f_{mil}\gamma^{jl})$. 

We thus have the four equations (\ref{bctEC}), (\ref{bcxEC}), (\ref{bcyEC}) and
(\ref{bczEC}) to be considered as potential boundary conditions for the EC
equations. Which ones are identically satisfied  by virtue of the EC equations
and the initial values, and which ones remain as boundary conditions is 
determined by the characteristic fields of the EC equations.  

The EC equations are symmetric hyperbolic (and thus strongly hyperbolic as
well~\cite{kreissbook}). With respect to the unit
vector $\xi^i \equiv \gamma^{xi}/\sqrt{\gamma^{xx}}$ which is
normal to the boundary $x=x_0$ for the region $x\le x_0$ there are 18 ``static'' characteristic fields (the six $\gamma_{ij}$, the six
$f^y{}_i{}^j$ and the six $f^z{}_i{}^j$) and 12 characteristic fields traveling
at the speed of light, of which six are incoming:
\begin{equation}
{}^-U^j_i \equiv K^j_i -\frac{f^x{}_i{}^j}{\sqrt{\gamma^{xx}}}
\end{equation}

\noindent and six are outgoing:
\begin{equation}
{}^+U^j_i \equiv K^j_i + \frac{f^x{}_i{}^j}{\sqrt{\gamma^{xx}}}
\end{equation}

\noindent With regards to boundary values, those of the static and outgoing
fields are completely determined by the initial values. Because the time
derivatives of the incoming fields ${}^-U^y_z, {}^-U^x_x$ and the combination
${}^-U^y_y-{}^-U^z_z$ do not occur in any of the four boundary equations
(\ref{bctEC}), (\ref{bcxEC}), (\ref{bcyEC}) or (\ref{bczEC}), the values of
these three incoming fields on the boundary are completely
\textit{arbitrary\/}.  However, the time derivatives of ${}^-U^x_y$ and
${}^-U^x_z$  occur in (\ref{bcyEC}) and (\ref{bczEC}), respectively, so the
boundary values  of ${}^-U^x_y$ and ${}^-U^x_z$ are determined by (\ref{bcyEC})
and (\ref{bczEC}) in terms of outgoing and static characteristic fields, as
well as initial values.  Finally, the time derivatives of the combination
${}^-U^y_y+{}^-U^z_z$ appear in (\ref{bctEC}) and in (\ref{bcxEC}), as well as
those of the corresponding ougoing counterpart ${}^+U^y_y+{}^+U^z_z$. This
means that the two equations (\ref{bctEC}) and (\ref{bcxEC}) are equivalent to
one boundary prescription for an incoming field (${}^-U^y_y+{}^-U^z_z$) and one
condition on an outgoing field (${}^+U^y_y+{}^+U^z_z$) which must be identically
satisfied. 

Thus, in the EC formulation of the Einstein equations, \textit{three\/} of the
four equations $G_{ab}e^b=0$ are non trivial, as opposed to one in the ADM
case.


\section{Relation to constraint propagation in the EC formulation with
vanishing shift\label{sec:4}}

The question that arises in the EC case is how the three non-trivial boundary
conditions in the set $G_{ab}e^b=0$ relate to the constraints.  In analogy with
the ADM case of Section~\ref{sec:2}, we anticipate that in the EC case there
will be (at most) three constraints that will be ``non-preserved'' at the
boundary, and that they will be related to the three non-trivial components of 
$G_{ab}e^b=0$ by linear combinations with the evolution equations.  To prove
this would be far from trivial, in principle, but can indeed be done explicitly
by using the ADM case as a guide, as we show in the following. 

We start by looking at the auxiliary system of propagation of the constraints.
The reader should note, in the first place, that the addition of the 18
first-order constraints ${\cal C}_{kij}$ automatically raises the differential
order of the system of  propagation equations for the whole set of 22
constraints to second order: the evolution equations for the set $({\cal C},
{\cal C}_i, {\cal C}_{kij})$ implied by the EC equations involve second
space-derivatives of ${\cal C}_{kij}$. We have explicitly:
\begin{subequations}\label{constpropEC}
\begin{eqnarray}
\dot{\cal C}   &=&  \alpha \partial^i{\cal C}_i + \ldots\\
\dot{\cal C}_i &=& \frac12 \alpha\partial^k\left(
		   \partial_i{\cal C}_{kl}{}^l
		  -\partial_l{\cal C}_{ki}{}^l
   		  +\partial_k{\cal C}_{li}{}^l
 		  -\partial_k{\cal C}_{il}{}^l\right)\nonumber\\ &&
		  -\alpha \partial_i{\cal C}+ \ldots\\
\dot{\cal C}_{kij} &=& \ldots
\end{eqnarray}
\end{subequations}

\noindent where $\ldots$ denote undifferentiated terms. 
The system of auxiliary evolution equations (\ref{constpropEC}) for the 22
constraints is not manifestly well posed because of the presence of
second-derivatives in the right-hand side (in fact, some times this indicates
ill-posedness~\cite{bswave}). However, it can be reduced to first
differential order in the usual manner, by adding new ``constraint'' variables
that are space-derivatives of the constraints. In this case we can do with
\begin{equation}
{\cal C}_{lkij} \equiv \frac12 \left( \partial_l{\cal C}_{kij}
				     -\partial_k{\cal C}_{lij}\right),
\end{equation}

\noindent which enlarges the system (\ref{constpropEC}) to 40 variables in
all, and casts it in the following form:
\begin{subequations}\label{constpropECfirst}
\begin{eqnarray}
\dot{\cal C}   &=&  \alpha \partial^i{\cal C}_i + \ldots\\
\dot{\cal C}_i &=& -\alpha \left(\partial_i{\cal C}
		   +\partial^k{\cal C}_{lki}{}^l
		   +\partial^k{\cal C}_{kil}{}^l\right)
   		   + \ldots\\
\dot{\cal C}_{kij} &=& \ldots\\
\dot{\cal C}_{lkij}&=& \alpha\left(\gamma_{ki}\partial_l{\cal C}_j
		      +\gamma_{kj}\partial_l{\cal C}_i
		      -\gamma_{li}\partial_k{\cal C}_j
	              -\gamma_{lj}\partial_k{\cal C}_i\right) +\dots\nonumber\\
\end{eqnarray}
\end{subequations} 

\noindent We digress slightly now from the main point to point out that, if one
chooses to do so, the new ``constraint'' variables ${\cal C}_{lkij}$ can be
expressed in terms of the fundamental variables of the EC system, in which case
they read:
\begin{eqnarray}\label{Clkij}
{\cal C}_{lkij} &=& \partial_l(f_{kij} -2\gamma^{nm}(f_{nm(i}\gamma_{j)k}
	-\gamma_{k(i}f_{j)nm})) \nonumber\\ &&
	- \partial_k(f_{lij} -2\gamma^{nm}(f_{nm(i}\gamma_{j)l}
	-\gamma_{l(i}f_{j)nm})),\;\;\;\;\;
\end{eqnarray}

\noindent and turn out to be identical to the ``integrability conditions''
$0=1/2(\partial_l\gamma_{ij,k}-\partial_k\gamma_{ij,l})$. This is usually the
way that such ``constraint quantities'' necessary to reduce the propagation of
the constraints to first differential order are introduced by most authors,
starting with Stewart in \cite{stewart98}. They are not to be interpreted as
additional constraints to impose on the initial data, though. In fact, ${\cal
C}_{lkij} =0$ holds identically for any initial data satisfying ${\cal C}_{kij}
=0$. They are not to be used as substitutes for ${\cal C}_{kij}=0$ either,
because the vanishing of ${\cal C}_{kij}$ does not follow from the vanishing
of ${\cal C}_{lkij}$. Indeed, if  ${\cal C}_{lkij}=0$ is satisfied by $\gamma_{ij}$
and $f_{kij}$, then there exists a field $h_{ij}$ such that  $2f_{kij}
-4\gamma^{lm}(f_{lm(i}\gamma_{j)k} -\gamma_{k(i}f_{j)lm})=\partial_k h_{ij}$,
but it does not follow that $h_{ij} = \gamma_{ij}$. The introduction of the
additional ``constraints'' ${\cal C}_{lkij}=0$ is mostly a formal procedure to
study the evolution of the constraints as functions of the point --not through
the fundamental variables. In fact, one does not need to include all of  ${\cal
C}_{lkij}$ as additional variables in order to reduce the propagation of the
constraints to first differential order, since one can see that only the
contractions ${\cal C}_{lki}{}^l$ and ${\cal C}_{kil}{}^l$ appear in the
equations, and these are 12 variables, so at least six of the 18 new
variables ${\cal C}_{lkij}$ are entirely redundant.   

The immediate benefit of introducing the first-order constraint variables is
that the system (\ref{constpropECfirst}) is well posed in the sense that it is
strongly hyperbolic with characteristic speeds of 0 (multiplicity 34),
$+\alpha$ (multiplicity 3) and $-\alpha$ (multiplicity 3). This means that,
with respect to the (outer) boundary at $x=x_0$, three characteristic
constraint variables are incoming, three are outgoing and all the others are
``static''. The six non-static characteristic constraints, which we denote by
${}^\pm{\cal Z}_i$ (with $+$ for outgoing and $-$ for incoming), are
explicitly:
\begin{subequations}\label{Z's}
\begin{eqnarray}
{}^\pm {\cal Z}_x &=&
	{\cal C}_x \pm \frac{1}{\sqrt{\gamma^{xx}}}\left( {\cal C} +
	{\cal C}^{kx}{}_{xk}+{\cal C}^x{}_{xk}{}^k\right)\\
{}^\pm {\cal Z}_y &=&
	{\cal C}_y \pm \frac{1}{\sqrt{\gamma^{xx}}}
	\left({\cal C}^{kx}{}_{yk}+{\cal C}^x{}_{yk}{}^k\right)\\
{}^\pm {\cal Z}_z &=&
	{\cal C}_z \pm \frac{1}{\sqrt{\gamma^{xx}}}
	\left({\cal C}^{kx}{}_{zk}+{\cal C}^x{}_{zk}{}^k\right)
\end{eqnarray}
\end{subequations}
 
\noindent Because three of the characteristic constraints are incoming  at the
boundary, even if they are set to zero initially they will not be vanishing at
the boundary by virtue of the evolution equations. In fact, the three incoming
constraints must be prescribed at the boundary, either arbitrarily, or as
functions of the outgoing constraints. That is: the problem of the propagation
of the constraints, Eq.~(\ref{constpropECfirst}), requires three boundary
conditions.  Thus we have proven the
first part of our claim: there are three constraints that are ``non-preserved''
at the boundary, in the same number as nontrivial boundary conditions for the
EC equations,  as anticipated.  

The second part of the claim is to prove that the incoming constraints are
related to the equations $G_a^x=0$ by terms that are proportional to the
evolution equations. The size of the problem as regards the number of variables
makes this practically impossible unless one had a good guess as to what the
linear combinations ought to be. Guided by the ADM case, we may
expect that, through the evolution equations, $G^x_x$ will be related to ${\cal
C}$ and $G^x_t$ will be related to ${\cal C}^x$, except for terms involving the
new constraints ${\cal C}_{kij}$ or ${\cal C}_{lkij}$. We also expect that
$G^x_y$ and $G^x_z$ will only be related to the new constraints, but not the
hamiltonian nor vector constraints. Our ansatz is explicitly as follows:
\begin{subequations}\label{ansatz}
\begin{eqnarray}
G^x_t&\sim& \alpha{\cal C}^x \label{ansatza}\\
G^x_y&\sim& {\cal C}^{kx}{}_{yk} + {\cal C}^x{}_{yk}{}^k \label{ansatzb}\\
G^x_z&\sim& {\cal C}^{kx}{}_{zk} + {\cal C}^x{}_{zk}{}^k  \label{ansatzc}\\
G^x_x&\sim& {\cal C}+{\cal C}^{kx}{}_{xk}+{\cal C}^x{}_{xk}{}^k\label{ansatzd}
\end{eqnarray} 
\end{subequations}

\noindent which is equivalent to (\ref{bctoconst}) except for terms whose
occurrence is suggested by the combinations of constraints that can be
expressed directly in terms of linear combinations of the characteristic
constraints ${}^\pm{\cal Z}_i$.

The ansatz may be verified (or disproven) simply by comparing the right hand
sides of equations (\ref{bctEC}),  (\ref{bcyEC}), (\ref{bczEC}) and
(\ref{bcxEC}) with (\ref{ansatz}) term by term under the assumption that every
occurrence of a time derivative in (\ref{bctEC}),  (\ref{bcyEC}), (\ref{bczEC})
or (\ref{bcxEC}) must be substituted in terms of space-derivatives by means of
the evolution equations, that is:
\begin{subequations}\label{shortevol}
\begin{eqnarray}
\dot{K}_i^j &=& -\alpha \partial_k f^k{}_i{}^j +\dots\label{shortevola}\\
\dot{f}^k{}_i{}^j&=& -\alpha \partial_kK_i^j +\dots\label{shortevolb}
\end{eqnarray}
\end{subequations}

\noindent In particular, using (\ref{shortevola}) to eliminate $\dot{K}^x_y$,
for (\ref{bcyEC}) we have
\begin{equation}
G^x_y\sim \partial_lf^x{}_y{}^l +\partial_y(2f^k{}_k{}^x-3f^x{}_k{}^k )+\ldots
\end{equation}

\noindent On the other hand, \textit{by definition} from (\ref{Clkij}) we have
\begin{equation}
{\cal C}^{kx}{}_{yk}+{\cal C}^x{}_{yk}{}^k
= \partial_lf^x{}_y{}^l +\partial_y(2f^k{}_k{}^x-3f^x{}_k{}^k ) +\ldots
\end{equation}

\noindent Thus (\ref{ansatzb}) is \textit{verified}. By the same argument
substituting $y$ with $z$, (\ref{ansatzc}) is similarly verified. 

Next, using (\ref{shortevola}) to eliminate $\dot{K}-\dot{K}^x_x 
(=\dot{K}^y_y+\dot{K}^z_z)$, for (\ref{bcxEC}) we obtain:
\begin{eqnarray}
G^x_x&\sim& -\partial_x\left(f^x{}_y{}^y +f^x{}_z{}^z\right)\nonumber\\&&
	  +\partial_y\left(2f^y{}_x{}^x-f^x{}_x{}^y+2f^y{}_z{}^z
	  -2f^z{}_z{}^y\right) \nonumber\\ &&
	  +\partial_z\left(2f^z{}_x{}^x-f^x{}_x{}^z+2f^z{}_y{}^y
	  -2f^y{}_y{}^z\right) +\dots\;\;\;
\end{eqnarray}

\noindent On the other hand, directly by the definition (\ref{Clkij}) 
and the expression (\ref{CEC}) for the hamiltonian constraint we have
\begin{eqnarray}
\lefteqn{{\cal C}+{\cal C}^{kx}{}_{xk}+{\cal C}^x{}_{xk}{}^k } &&\nonumber\\
&=&-\partial_x\left(f^x{}_y{}^y +f^x{}_z{}^z\right)\nonumber\\&&
	  +\partial_y\left(2f^y{}_x{}^x-f^x{}_x{}^y+2f^y{}_z{}^z
	  -2f^z{}_z{}^y\right) \nonumber\\ &&
	  +\partial_z\left(2f^z{}_x{}^x-f^x{}_x{}^z+2f^z{}_y{}^y
	  -2f^y{}_y{}^z\right) +\dots\;\;\;
\end{eqnarray}

\noindent Thus also (\ref{ansatzd}) is verified. 

Finally, using (\ref{shortevolb}) to eliminate the time derivatives of
$f^k{}_i{}^j$ in terms of space derivatives of $K^j_i$, Eq.~(\ref{bctEC}) reads
\begin{equation}
G^x_t \sim -\alpha(\partial^x K - \partial^k K^x_k) +\ldots
\end{equation}

\noindent which is manifestly equal to $\alpha {\cal C}^x$, thus verifying
(\ref{ansatza}). 

If we represent the evolution equations (\ref{KdotEC}) and (\ref{fdotEC}) in
the form $\tilde{\cal E}_{ij}=0$ and $\tilde{\cal E}_{kij}=0$, respectively, by simply
transferring all the terms in the right-hand side to the left, what we have
shown is that the following relationships between the projection of the
Einstein tensor, the constraints and the evolution equations hold in the EC
formulation:
\begin{subequations}\label{bcrel}
\begin{eqnarray}
G^x_t&=& \alpha{\cal C}^x +\tilde{\cal E}^x{}_k{}^k - \tilde{\cal E}^k{}_k{}^x\\
G^x_y&=& {\cal C}^{kx}{}_{yk} + {\cal C}^x{}_{yk}{}^k 
- \alpha^{-1} \tilde{\cal E}^x_y \\
G^x_z&=& {\cal C}^{kx}{}_{zk} + {\cal C}^x{}_{zk}{}^k 
- \alpha^{-1} \tilde{\cal E}^x_z \\
G^x_x&=& {\cal C}+{\cal C}^{kx}{}_{xk}+{\cal C}^x{}_{xk}{}^k
+ \alpha^{-1}\tilde{\cal E}^k_k - \alpha^{-1}\tilde{\cal E}^x_x 
\end{eqnarray}
\end{subequations}

\noindent It may be objected that the proof is not complete because the
undifferentiated terms of the equations have not been shown explicitly to be
the same on both sides. We think that it is quite
clear that a complete proof in that sense may not be feasible, but yet the
consistency of the principal terms and the geometrical foundation of all terms
on both sides, taken together, give a very strong indication that the equality
will hold term by term. 

The relationship to the characteristic constraints is found by ``inverting''
(\ref{Z's}) in order to have the fundamental constraints in terms of the
characteristic constraints:
\begin{subequations}
\begin{eqnarray}
{\cal C}_i &=& \frac12\left(  {}^+{\cal Z}_i +{}^-{\cal Z}_i\right)\\
{\cal C}^{kx}{}_{yk}+{\cal C}^x{}_{yk}{}^k &=&
\frac{\sqrt{\gamma^{xx}}}{2}\left(  {}^+{\cal Z}_y -{}^-{\cal Z}_y\right)\\
{\cal C}^{kx}{}_{zk}+{\cal C}^x{}_{zk}{}^k &=&
\frac{\sqrt{\gamma^{xx}}}{2}\left(  {}^+{\cal Z}_z -{}^-{\cal Z}_z\right)\\
{\cal C} +
{\cal C}^{kx}{}_{xk}+{\cal C}^x{}_{xk}{}^k &=& 
\frac{\sqrt{\gamma^{xx}}}{2}\left(  {}^+{\cal Z}_x -{}^-{\cal Z}_x\right)
\end{eqnarray}
\end{subequations}
 
\noindent Thus, up to terms proportional to the evolution equations, imposing
$G^x_a=0$ is equivalent to imposing
\begin{subequations}\label{bcZ's}
\begin{eqnarray}
\gamma^{xi}({}^+{\cal Z}_i +{}^-{\cal Z}_i) &=& 0\label{bcZ'sa}\\
{}^+{\cal Z}_y -{}^-{\cal Z}_y &=& 0\label{bcZ'sb}\\
{}^+{\cal Z}_z -{}^-{\cal Z}_z &=& 0\label{bcZ'sc}\\
{}^+{\cal Z}_x -{}^-{\cal Z}_x &=& 0\label{bcZ'sd}
\end{eqnarray}
\end{subequations}

\noindent From our discussion in Section~\ref{sec:3} it follows that
(\ref{bcZ'sb}) and (\ref{bcZ'sc}) need to be imposed, and then either
(\ref{bcZ'sa}) or (\ref{bcZ'sd}) or, in fact, any linear combination of them,
as the third boundary condition for the EC equations. Remarkably, this is
entirely consistent with the constraint propagation problem
(\ref{constpropECfirst}), since (\ref{bcZ'sb}), (\ref{bcZ'sc}) plus
\textit{one\/} linear combination of (\ref{bcZ'sa}) and (\ref{bcZ'sd}) turn out
to be an admissible complete set of boundary conditions for the three incoming
constraints in terms of the three outgoing constraints. By admissible boundary
conditions we mean, for instance, maximally dissipative boundary conditions,
which, as is known~\cite{kreissbook}, for homogeneous problems like this one, essentially take
the form of ${}^-U_i = L_i^j{}^+U_j$ with a rather general (bounded) matrix $L_i^j$.

In fact, the reader will have no difficulty to recognize in (\ref{bcZ'sa}), 
(\ref{bcZ'sb}) and (\ref{bcZ'sc}) the ``constraint-preserving'' boundary conditions
of the Neumann type of~\cite{calabrese3d} (with $\eta=4$), where boundary conditions
arising from constraint propagation are discussed in a linearized setting (and
furthermore restricting $\eta$ to the interval $0<\eta<2$.) This means that the set
of three boundary equations
\begin{subequations}
\begin{eqnarray*}
G^x_t &=& 0,\\
G^x_y &=& 0,\\
G^x_z &=& 0,
\end{eqnarray*}
\end{subequations}

\noindent for the EC formulation --with $G^x_t,G^x_y,G^x_z$ given by (\ref{det}),
(\ref{Kxy}) and (\ref{Kxz}) written in terms of the variables $f_{kij}$ in such a
way that the principal terms occur as in (\ref{bctEC}), (\ref{bcyEC}) and
(\ref{bczEC}) respectively-- constitute the \textit{exact\/} (non-linear) form of
the so called constraint-preserving boundary conditions of the Neumann type that
one would write for the  case of the Einstein-Christoffel formulation by following
the prescription in~\cite{calabrese3d}.  This result generalizes to three dimensions
the prediction that we advanced in \cite{0302032}, where the constraint preserving
boundary conditions were found to be identical to the projections of the Einstein
tensor normal to the boundary for the EC equations with the restriction of spherical
symmetry. 

What needs to be made clear, however, is the question of why should the
equations $G^x_a=0$ imposed on the boundary enforce the constraints in the
interior, given that they are only equivalent to the constraints at places
where the evolution equations are satisfied. As a matter of fact, the
evolution equations are \textit{not\/} imposed on the boundary (that's why
one needs boundary prescriptions). Therefore it is not true that the
constraints are being enforced on the boundary. Still, they are enforced in
the interior, which is the goal. The argument is the following.  The
vanishing of $G^x_a$ on the boundary is, in a sense,``carried'' by the
incoming fields to the interior, where the evolution equations are actually
satisfied. In the interior, then, one has both $G^x_a=0$ and ${\cal
E}_{ij}={\cal E}_{ijk}=0$. Therefore, by (\ref{bcrel}), the constraints
related to $G^x_a$ are enforced in the interior. Thus, by imposing $G^x_a=0$
on the boundary, one enforces the constraints that would otherwise be
violated outside of the domain of dependence of the initial surface.

\section{Stability considerations}\label{sec:5}

The fact that three linearly independent combinations of $G^x_a=0$ except
$G^x_t+\gamma^{xi}G^x_i=0$ yield necessary and sufficient boundary conditions on the
incoming fundamental fields of the EC equations has importance in its own right:
they ensure the uniqueness of solutions to the initial-boundary-value problem.
Additionally, the fact that such boundary conditions are related to the non-static
characteristic constraints by linear combinations with the evolution equations is
entirely equivalent to a procedure of ``trading'' transverse space derivatives by
time derivatives using the evolution equations, which is the procedure that has been
used in \cite{calabrese3d} to obtain what the authors refer to as
constraint-preserving boundary conditions.  Which of the two meanings (linear
combinations with evolution versus trading derivatives) to use is only a matter of
taste. We (as the authors of \cite{calabrese3d} apparently do) interpret this fact
as an indication that the boundary conditions of the evolution problem are
equivalent to imposing vanishing values of the constraints on the boundary (up to
linear combinations of the constraints among themselves).

The reader with an interest in numerical applications may be concerned with the
question of whether any or all sets of Einstein boundary conditions (by which we
mean three linearly independent combinations of $G^x_a=0$ except
$G^x_t+\gamma^{xi}G^x_i=0$) preserve the well-posedness of the \textit{initial}
value problem represented by the EC equations. More precisely, with the question of 
which sets of Einstein boundary conditions ensure that the solution at the final
time depends continuously on the initial data.  This question has not been answered
in the previous sections, as it lies beyond the scope of the present work, requiring
a calculation by means of either energy methods (in the best case) or Laplace
transform methods (in the most likely case), as described in \cite{kreissbook}.

However, preliminary results of relevance to this question already exist in the
recent literature, and it is appropriate to mention them and their immediate
consequences.  In \cite{calabresesarbach} the authors undertake a Laplace-transform
type of study of the well-posedness of several boundary prescriptions including some
choices of the Einstein boundary conditions for the standard EC system, as in the
present article (as opposed to ``generalized'' EC as in \cite{calabrese3d}).  In
Section IV of \cite{calabresesarbach} the authors conclude that, in the
linearization around Minkowski spacetime, the choice of three boundary conditions
as  $G_{xt}=G_{xy}=G_{xz}=0$ is in fact well posed for the case of standard EC as in
the present article. Since the linearization of $G^x_t=G^x_y=G^x_z=0$ around
Minkowski space coincides with $G_{xt}=G_{xy}=G_{xz}=0$, this in fact proves that,
in the linearization around flat space, the three Einstein boundary conditions 
$G^x_t=G^x_y=G^x_z=0$ constitute a well-posed set of boundary conditions for the
(standard) EC formulation. 

Moreover, by the present paper, these three Einstein boundary conditions represent
the well-posed Neumann boundary conditions of the ``constraint-preserving'' scheme
of \cite{calabrese3d} if one were to extend those authors' argument and lexicon to
the case $\eta=4$. In the constraint-preserving scheme of \cite{calabrese3d} the
authors did not allow the parameter of the generalized EC system to take the value
$\eta=4$ required for the standard EC system, for the reason that the energy method
of \cite{calabrese3d} works well only for symmetric hyperbolic formulations with
symmetric hyperbolic constraint propagation, which is not the case for $\eta\ge 2$. 
As a result, in \cite{calabrese3d} the authors prove that $\eta<2$ (necessary for
symmetric hyperbolic constraint propagation) is sufficient for well-posed
constraint-preserving boundary conditions, but they do not prove that $\eta<2$ is
necessary.  On our part, by combining the results of \cite{calabresesarbach} (in
which the Laplace-transform method is used rather than energy methods) with our
current article as indicated, we are indeed demonstrating that $\eta<2$ is, in fact,
\textit{not} necessary in order to have well-posed constraint-preserving boundary
conditions of the Neumann type (this fact is actually indicated implicitly --but
unambiguously-- in \cite{calabresesarbach}). This also disproves, by counterexample,
the erroneous belief that the propagation of the constraints needs to be symmetric
hyperbolic (not just strongly hyperbolic) in order for such well-posed constraint
preserving boundary conditions to exist. But more interestingly, we are providing
the principal terms of the exact nonlinear boundary conditions, the non-principal
ones being precisely indicated by the exact expression of the components of the
Einstein tensor, namely by full substitution of the fundamental variables in
Eqs.~(\ref{bcs}). 

It also needs to be pointed out to the readers that in the same paper
\cite{calabresesarbach} the authors show that imposing the linear combination
$G_{xx}-G_{tt}=0$ in addition to  $G_{xy}=G_{xz}=0$ leads to an ill-posed problem in
the linearization around Minkowski space.  This is an indication that
$G^x_t-\gamma^{xi}G^x_i=0$ used with $G^x_y=G^x_z=0$ may lead to an ill-posed
initial-boundary-value problem in the nonlinear case as well, though an actual proof
for the nonlinear case is still lacking. As terminology goes, readers should be made
aware that the authors of \cite{calabresesarbach} inappropriately associate the
concept of Einstein boundary conditions exclusively with this set, without
acknowledging that, in the way we have used it, the term refers to all linear
combinations of $G^x_a=0$, including the well-posed ones, and including those that,
as we show, are equivalent to the ``constraint-preserving'' boundary conditions of
\cite{calabrese3d} in the linearization.

\section{Concluding remarks}\label{sec:6}

What has been shown is that the vanishing of the projection of the Einstein
tensor normal to the boundary acts as boundary conditions for the EC equations
that automatically enforce the constraints that are otherwise not propagated, and
that the set $G^x_t=G^x_y=G^x_z=0$, in particular, leads to a well-posed
initial-boundary-value problem ideally suited for numerical evolution. 

The reader can see that the concept of Einstein boundary conditions can be applied
to \textit{any} formulation of the initial-boundary value problem of the Einstein
equations. The choice of the EC formulation in this work is merely done for
convenience as the clearest illustration. We think it is reasonable to postulate
that in the case of \textit{any\/} strongly hyperbolic formulation of the Einstein
equations the Einstein boundary conditions will include a set that will be
equivalent to constraint-preserving boundary conditions of the Neumann type. But
most numerical simulations today are currently done using formulations that are not
strongly hyperbolic, in which case the question of well-posedness becomes
irrelevant, yet the problem of identifying useful and consistent boundary
conditions remains. For such formulations, the Einstein boundary conditions
eliminate some guesswork on boundary values and may help control the constraint
violations. 

Given the explicit expressions (\ref{bcs}), the Einstein boundary conditions
$G_{ab}e^b=0$ (up to linear combinations as indicated throughout) for any 3+1
formulation of the Einstein equations are found by simply expressing (\ref{bcs}) in
the chosen representation of fundamental variables. The only subtlety that must be
taken care of is that the final form of $G_{ab}e^b=0$ in the chosen variables must
contain no derivatives across the boundary of the same order as the evolution
equations. That such a form exists is guaranteed by the Bianchi identities. In all
cases, the procedure to obtain the correct form of the Einstein boundary conditions
in the chosen variables parallels what is done in the current work for the case of
the EC formulation. In particular, with perhaps very little effort our method can
almost certainly be used to show (or disprove) our claim that the
constraint-preserving boundary conditions of the Neumann type for the generalized EC
systems as in \cite{calabrese3d} are, indeed, Einstein boundary conditions as well.

\begin{acknowledgments}
We are indebted to Carsten Gundlach for kindly pointing out a significant error
in an earlier version. This work was supported by the NSF under grants No.
PHY-0070624 and PHY-0244752 to Duquesne University, and No. PHY-0135390 
to Carnegie Mellon University.
\end{acknowledgments}


\end{document}